\documentclass{ws-ijmpe}

\begin{document}

\markboth{N. Schunck, P. Olbratowski, J. Dudek, and J. Dobaczewski}{Rotation of
Tetrahedral Nuclei in the Cranking Model}

\catchline{}{}{}{}{}
%
%%%%%%%%%%%%%%%%%%%%%%%%%%%%%%%%%%%%%%%%%%%%%%%%%%%%%%%%%%%%%%%%%%%%

\title{ROTATION OF TETRAHEDRAL NUCLEI IN THE CRANKING MODEL}

\author{\footnotesize N. SCHUNCK$^{1,2}$, P. OLBRATOWSKI$^{2}$,
J. DUDEK$^{3}$, and J. DOBACZEWSKI$^{2}$}

\address{\it
(1) Departamento de Fisica Teorica, Universidad Autonoma de Madrid, \\
Cantoblanco, 28 049 Madrid, Spain\\
(2) Institute of Theoretical Physics Warsaw University, Ho\.za 69, PL-00681 Warsaw, Poland\\
(3) Institut de Recherches Subatomiques, IN$_2$P$_3$-CNRS/Universit\'e Louis Pasteur \\
    F-67037 Strasbourg Cedex 2, France\\
}

\maketitle

\begin{history}
\received{(received date)}
\revised{(revised date)}
%\accepted{(Day Month Year)}
%\comby{(xxxxxxxxxx)}
\end{history}

\begin{abstract}
The three-dimensional cranking model is used to investigate the microscopic
aspects of the rotation of nuclei with the tetrahedral symmetry. Two classes of
rotation axes are studied corresponding to two different discrete symmetries of
the rotating hamiltonian. Self-consistent Hartree-Fock-Bogoliubov calculations 
show that the tetrahedral minimum remains remarkably stable until the first 
single-particle crossing. 
\end{abstract}

\section{Introduction}

The prediction that atomic nuclei can possess stable low-lying configurations
with the tetrahedral symmetry has been confirmed by several independent
calculations using a variety of nuclear 
mean-fields\cite{JDu02,RapCom,TYM98,przemek}. However, no experimental evidence 
has been reported so far, which can be put down to the lack of understanding of 
the excitation mechanisms in such exotic systems. We investigate here the
collective rotation, since the deep minima and large barriers reported e.g.\
in Ref.\cite{RapCom} suggest tetrahedral nuclei may be amenable to sustain
rotational bands. This article is a sequel to previous works where the
structure and moments of inertia of tetrahedral collective bands were
analyzed\cite{prl}, and the consequences of the collective rotation in terms
of symmetries and quantum numbers were discussed\cite{tdrot}. To complement
these studies, we wish to focus here on the stability of the tetrahedral
minimum as a function of the rotational frequency, which is one of the
pre-requisites to an eventual observation of rotational bands.

Today there exists several articles in the literature about the nuclear
tetrahedral symmetry (associated to the point group of symmetry $T_{d}^{D}$)
and we refer the reader interested in a general introduction to this subject
to e.g.\ Ref.\cite{JDu02a}. For the purpose of the present study, it suffices to
recall that a nucleus with a non-axial $\alpha_{32}$ octupole deformation (in 
the standard expansion of the nuclear radius on the basis of spherical 
harmonics, with deformation parameters $\alpha_{\lambda\mu}$) has all the 
symmetries of a regular tetrahedron. Although other realizations of the group 
$T_{d}^{D}$, involving higher-order multipole terms with $\lambda\geq 7$, also 
appear to generate stable minima in the potential energy landscape\cite{dudek}, 
they will not be considered in this article.

The cranking model was the subject of several comprehensive review articles and
we refer to Refs.\cite{MdV83,Sat05} for a general discussion of its main 
features. We just recall here how the model should be adapted in the 
3-dimensional case: the rotation is described with the help of 3 Lagrange 
multipliers $\vec{\omega}\equiv (\omega_x, \omega_y, \omega_z)$ which are 
interpreted as the classical rotational frequencies along the x-, y- and z-axis 
of the body-fixed frame respectively. Equivalently, one can choose the 
spherical representation in which the rotational frequency vector 
$\vec{\omega}$ is parameterized by $(\omega, \theta, \varphi)$.

It is known that a collective rotation of a nucleus with prolate or oblate
deformation takes place about an axis perpendicular to the symmetry axis. 
However, in the case of a nuclear shape with the $T_{d}^{D}$ symmetry, the 
quadrupole deformation is equal to zero, and consequently no simple criterion 
exists to determine the sorientation of the rotation axis.

A possible procedure to find optimum rotation axis consists in computing the 
total energy as a function of the orientation of the rotation axis, 
characterized by the two angles $(\theta,\varphi)$, for different rotational 
frequencies $\omega$, as was done in Refs.\cite{prl,tdrot}. One thus obtains 
two-dimensional maps whose minima signal the energetically-favored axes of 
rotation. These calculations were performed using a macroscopic-microscopic 
technique, in which the total energy is the sum of a liquid-drop contribution 
parameterized as in Ref.\cite{lsd} and shell-correction extracted from a 
Woods-Saxon potential with the form defined in Ref.\cite{universal}. The 
results\cite{prl,tdrot} suggest that at low rotational frequencies (up to 
$\omega \sim 0.3 $ MeV/$\hbar$), no particular axis of rotation is favored, 
while at higher frequencies, several well-defined minima emerge. However, these 
calculations assumed a fixed tetrahedral deformation and did not include 
pairing correlations: therefore, in the present study we want to see whether
the tetrahedral minimum survives the increase of angular momentum.

\section{Stability of the tetrahedral minimum}

In order to study the stability of the tetrahedral minimum and to treat on the
same footing the response of the nuclear system to rotation and pairing
correlations, we performed the Hartree-Fock-Bogoliubov (HFB) calculations in
the current implementation of the code {\sc hfodd} (v2.17k)\cite{hfodd} for 
$^{110}$Zr, which shows\cite{RapCom} a pronounced tetrahedral minimum 
at $I = 0$. In our calculations we employed the SLy4\cite{sly4} Skyrme force
in the particle-hole channel and the density-dependent delta interaction in the 
particle-particle channel, whose intensity was fitted so as to reproduce the 
trend of experimental pairing gaps in the neutron-rich Zr isotopes, see 
Ref.\cite{przemek} for details.

The calculations were performed for two families of rotation axes: (i) the axis
passing through the middle of the edge of the tetrahedron, (ii) the axis passing
through the tip of the tetrahedron. They will be referred hereafter as to edge
and tip axes, respectively. These two axes are given a particular attention as
they are symmetry axes of the rotating tetrahedron; a 4-fold axis with inversion 
(edge) and a 3-fold axis (tip), i.e. these symmetries commute with the
cranking hamiltonian.

For the edge and tip axes, we performed the constrained HFB calculations.
In the case of the edge axis we requested that the discrete antilinear 
$T$-simplexes, $S_{x}^{T}$ and $S_{y}^{T}$, be conserved. (We refer to Ref.
\cite{hfodd} for a comprehensive review of the discrete symmetries implemented 
in the code {\sc hfodd}). This was equivalent to conserving the $z$-signature 
symmetry, $R_{z}$. They were supplemented by full symmetry-unconstrained 
calculations, which was the only option available for the rotation about the 
tip axis. In the latter case, we also used the option to readjust the 
orientation of the rotation axis self-consistently in the course of the 
iterations. We did not notice (i) any change in the orientation of $\vec{\omega}$ 
fixed at the beginning of the iterations, at least up to the first 
single-particle crossing, (ii) any significative difference in energy between 
the various orientations of $\vec{\omega}$ and between the symmetry-constrained 
and symmetry-unconstrained case. This confirms a spherical-like behaviour of 
the rotating tetrahedron as claimed in Ref.\cite{prl}.

\begin{figure}[th]
\vspace*{-0.2cm}
\centerline{\psfig{file=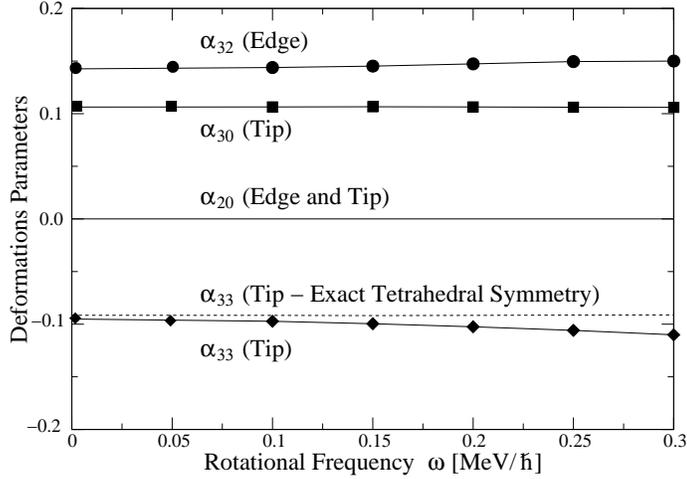,width=9.0cm}}
\caption{Evolution of the deformation parameters in the tetrahedral minimum of
$^{110}$Zr as function of the rotational frequency $\omega$ for the
edge- (circles) and tip-axes (squares and diamonds). For both classes of axes,
the quadrupole deformation remains equal to 0. For the tip-axis, the 
theoretical value of $\alpha_{33}$ in the exact $T_{d}^{D}$ limit is marked as 
the dashed line.}
\vspace*{-0.2cm}
\label{fig01}
\end{figure}

With the standard choice of the spherical harmonics for the calculation of
the multipole moments, the edge-axis coincides with the z-axis of the body-fixed
frame\cite{tdrot}. The deformations $\alpha_{\lambda\mu}$ extracted from the
multipole moments can be plotted directly, cf. Fig. \ref{fig01}. The curve with 
plain circles shows the "tetrahedral" deformation $\alpha_{32}$. It is worth
noticing that it is nearly constant as function of $\omega$, thus proving the 
resilience of the tetrahedral symmetry to the rotation.

In the case of the tip-axis, we chose to first perform a rotation of the 
nucleus at rest by the Euler angles $(\alpha,\beta,\gamma)$ in such a way that 
the new tip-axis coincides with the z-axis of the reference frame. This 
corresponds to an angle $\beta$ such that $\cos^{2}\beta = 1/3$, and  
$\alpha = \gamma = 0$. The multipole moments $Q_{\lambda\mu}$  
must be transformed according to the general relation:
\begin{equation}
Q'_{\lambda\mu} = \sum_{\mu'}
D^{*\lambda}_{\mu\mu'}(\alpha,\beta,\gamma)Q_{\lambda\mu}
\end{equation}
where the $D^{*\lambda}_{\mu\mu'}(\alpha,\beta,\gamma)$ are the Wigner matrices.
In our case, the initial deformation was characterized by $Q_{32} \neq 0$ and
all other moments were null. After the rotation, only $Q_{30}$ and $Q_{33}$
are different from zero, and they are related by:
$Q_{30}/Q_{33} = -\sqrt{10}/2$. We report in Fig. \ref{fig01} the equivalent
$\alpha_{30}$ and $\alpha_{33}$ deformations as function of the rotational 
frequency, together with the value of $\alpha_{33}$ in the exact symmetry limit.
The little deviation from the exact symmetry case seems to increase slightly
with $\omega$ although remaining very small.

In conclusion, we have investigated the rotation of tetrahedral nuclei using a
fully-microscopic 3-dimensional cranking model. The results of self-consistent
Skyrme Hartree-Fock Bogoliubov calculations show that the tetrahedral minimum
remains as function of the rotational frequency and that practically no
quadrupole polarization remains. The latter observation implies strongly 
hindered stretched-$E2$ transitions for the eventual rotational bands.

This work was supported in part by the exchange programme between
IN2P3 (France) and Polish Nuclear Physics Laboratories, by the Polish
Committee for Scientific Research (KBN) under Contract No.~1~P03B~059~27 
and by the Foundation for Polish Science (FNP).

\end{document}